\documentclass{emulateapj}
\usepackage{natbib}

\shortauthors{T. Hosokawa et al.}
\shorttitle{Supergiant Protostars}

\newcommand{\msunyr}{M_\odot~{\rm yr}^{-1}}
\newcommand{\mdot}{\dot{M}_*}
\newcommand{\ledd}{L_{\rm Edd}}

\begin{document}

\title{Rapidly Accreting Supergiant Protostars: \\
Embryos of Supermassive Black Holes?}
\author{Takashi Hosokawa\altaffilmark{1,2},
        Kazuyuki Omukai\altaffilmark{2},
        Harold W. Yorke\altaffilmark{1}}

\altaffiltext{1}{Jet Propulsion Laboratory, California Institute
of Technology, Pasadena CA 91109, USA; 
Takashi.Hosokawa@jpl.nasa.gov, hosokwtk@gmail.com}
\altaffiltext{2}{Department of Physics, Kyoto University, 
Kyoto 606-8502, Japan; omukai@tap.scphys.kyoto-u.ac.jp}

\begin{abstract}
Direct collapse of supermassive stars (SMSs) is a possible 
pathway for generating supermassive black holes in the early universe. 
It is expected that an SMS could form
via very rapid mass accretion with $\mdot \sim 0.1 - 1~\msunyr$
during the gravitational collapse of an atomic-cooling 
primordial gas cloud. 
In this paper we study how stars would evolve under such extreme
rapid mass accretion, focusing on the early evolution until the stellar 
mass reaches $10^3~M_\odot$.
To this end we numerically calculate the detailed interior structure
of accreting stars with primordial element abundances.
Our results show that for accretion rates higher than
$10^{-2}~\msunyr$, stellar evolution is qualitatively different from 
that expected at lower rates.
While accreting at these high rates
the star always has a radius exceeding $100~R_\odot$,
which increases monotonically with the stellar mass.
The mass-radius relation for stellar masses exceeding 
$\sim 100~M_\odot$ follows the same track with $R_* \propto 
M_*^{1/2}$ in all cases with accretion rates $\ga 10^{-2}~\msunyr$;
at a stellar mass of $10^3~M_\odot$ the radius is
$\simeq 7000~R_\odot$ ($\simeq 30$~AU).
With higher accretion rates the onset of hydrogen burning is 
shifted towards higher stellar masses. 
In particular, for accretion rates exceeding
$\mdot \gtrsim 0.1~\msunyr$, there is no significant hydrogen burning
even after $10^3~M_\odot$ have accreted onto the protostar.
Such ``supergiant'' protostars have effective temperatures as low as
$T_{\rm eff} \simeq 5000$~K throughout their evolution and because 
they hardly emit ionizing photons, they do not create an HII region or
significantly heat their immediate surroundings.
Thus, radiative feedback is unable to hinder the growth of 
rapidly accreting stars to 
masses in excess of $10^3~M_{\odot}$, as long as material is accreted
at rates $\mdot \gtrsim 10^{-2}~\msunyr$.
\end{abstract}


\section{Introduction}
\label{sec:intro}

Recent observations reveal that supermassive black holes
(SMBHs) exceeding $10^9~M_\odot$ already existed in the universe 
less than 1~Gyr
after the Big Bang \citep[e.g.,][]{Fan06, Mortlock11, Treister11}. 
The origins of such SMBHs must be intimately related to structure
formation in the early universe.
Some scenarios on the birth and growth of SMBHs postulate the 
existence of remnant BHs from Population III (Pop III) stars as 
their seeds \citep[e.g.,][]{MR01,Sch02}.
For several decades theoretical studies have predicted that 
the majority of Pop III stars were very massive, exceeding $100~M_\odot$
\citep[e.g.,][]{BL04}.
Pop III stars more massive than $300~M_\odot$ end their lives by
directly collapsing to form BHs \citep[e.g.,][]{HW02}.
If such a $\sim 100~M_\odot$ BH grows via continuous mass accretion
at the Eddington limited rate, its mass barely attains 
$10^9~M_\odot$ in 1~Gyr.

This scenario, however, has recently been challenged.
First, it is suspected that most Pop III stars were
much less massive than previously thought.
A circumstellar disk forming after the cloud's collapse easily
fragments due to gravitational instability and could produce multiple
protostars \citep[e.g.,][]{Machida08,Stacy10,Cl11}.
The final stellar masses would be reduced as the accreting
gas is shared by multiple stars.
Moreover, strong stellar UV light creates an HII region around
the protostar when the stellar mass exceeds a few $\times 10~M_\odot$.
The resulting feedback terminates the growth of Pop III protostars 
via mass accretion at a few $\times 10~M_\odot$ 
\citep[e.g.,][]{MT08,HOYY11,Stacy12}.
A large amount of gas would be expelled from the dark halo
due to the expansion of HII regions and the onset of
core-collapse supernovae
\citep[e.g.,][]{WAN04,Kitayama04,KY05}, 
which quenches the supply of gas to any remnant BH.
Even if a BH gets some gas supply, radiative feedback
from the BH accretion disk could regulate mass accretion
onto the BH-disk system \citep[e.g.,][]{Alv09,Mil09,Jeon11}. 

Another pathway for generating SMBHs is BH binary mergers.
However, this process could be also limited due to the
strong recoil resulting from gravitational wave
emission \citep[e.g.,][]{Campanelli07,Herrmann07}.
It is not straightforward that a  seed BH $\lesssim 100~M_\odot$ can grow 
to a $\sim 10^9~M_\odot$ SMBH within 1~Gyr of its birth.

An alternative possibility is that massive BHs exceeding
$10^5~M_\odot$ form directly
in some rare occasions in the primordial gas
\citep[e.g.,][]{BL03}. The primary cooling process in the primordial 
gas is line emission of molecular hydrogen.
However, the thermal evolution of a gravitationally collapsing 
cloud can change significantly, if this cooling
process is suppressed, for example, due to photodissociation
of molecules by strong background radiation 
\citep[][]{Omukai01,OH02,Shang10,IO11} 
or collisional 
dissociation in dense shocks \citep[][]{IO12}.
If the dark halo is sufficiently massive ($\ga 10^{8}M_{\odot}$), 
the baryonic gas can contract with atomic hydrogen cooling 
even without molecular hydrogen.
The collapse proceeds nearly isothermally at $\simeq 8000$~K. 
Without efficient molecular cooling fragmentation is suppressed
and single or binary protostars form within dense cloud cores of 
$\ga 10^5~M_\odot$ \citep{BL03,RH09}.
The protostar's mass is initially $\sim 10^{-2}~M_\odot$ but 
quickly increases via mass accretion.
The expected accretion rates are $0.1 - 1~\msunyr$, 
more than 100 times higher than the standard value
$\simeq 10^{-3}~\msunyr$ expected for Pop III star formation.
The stellar mass could reach $10^{5-6}~M_\odot$ in $\sim 1$~Myr
with this very rapid mass accretion.
General relativity predicts that such a supermassive star (SMS)
becomes unstable \citep[e.g.][]{Chandra64} and collapses to 
form a BH, which subsequently swallows most of 
the surrounding stellar material \citep[e.g.,][]{Shibata02}. 
Some authors are exploring a different picture, whereby only a central
part of the SMS collapses to form a $\sim 100~M_\odot$ BH and heat
input from the accreting BH inflates the outer envelope of the SMS
\citep[''quasi-star'',][]{Begelman06, Begelman08, Begelman10,
Ball11, Dotan11}.

However, we only have limited knowledge on how stars evolve
under such extreme conditions of rapid mass accretion.
\citet{Begelman10} predicts that, based on 
simple analytic arguments, such stars have a very different structure
from their main-sequence counterparts.
Stellar evolution at lower accretion rates 
$\mdot \lesssim 10^{-2}~\msunyr$ has been studied 
in detail by numerically solving the stellar interior structure
\citep[e.g.,][]{SPS86,OP01,OP03,Ohkubo09,HO09a,HYO10}.
\citet{OP01,OP03} showed that rapid mass accretion with 
$\mdot > 4 \times 10^{-3}~\msunyr$ causes the protostar's
abrupt expansion before its arrival to the Zero-Age Main 
Sequence (ZAMS).
Further comprehensive studies on stellar evolution with
rapid mass accretion are indispensable for considering 
their radiative feedback and observational signatures 
\citep[e.g.,][]{Johnson11}.

We present here our first results of this sort, whereby
as a first step, we study the early evolution up to a stellar mass 
of $10^3~M_\odot$. Our results show that rapid accretion with 
$\mdot \gtrsim 10^{-2}~\msunyr$ causes the star to 
bloats up like a red giant.
The stellar radius increases monotonically with stellar mass
and reaches $\simeq 7000~R_\odot (\simeq 30$~AU) at a
mass of $10^3~M_\odot$.
Unlike the cases with lower accretion rates previously 
studied, the mass-radius relation in this
phase is almost independent of the assumed accretion rate.
Such massive ``super-giant'' protostars could be the progenitors 
that eventually evolve to the observed SMBHs in the early universe.

The organization of this paper is as follows. 
First, we briefly review our numerical method 
and summarize the calculated cases in Section \ref{sec:model}.
We describe our results in Section \ref{sec:results};
we first focus on the fiducial case with $\dot{M} = 0.1~\msunyr$
and then examine effects of varying accretion rates and 
boundary conditions. 
Finally, summary and discussions are described 
in Section \ref{sec:sum}.

\begin{deluxetable*}{cccccc}
\tablecaption{
Cases Considered
\label{tb:md}
}
\tabletypesize{\scriptsize}
\tablehead{
\colhead{Case} &
\colhead{$\mdot$ ($\msunyr$)} &
\colhead{$M_{*,0}$ ($M_\odot$)} &
\colhead{$R_{*,0}$ ($R_\odot$)} &
\colhead{Notes} &
\colhead{References} 
}
\startdata
MD1e0  & $1.0$  & 2.5 & 298.4 & &  Sec.~\ref{ssec:vrate} \\
MD3e1  & $0.3$  & 2.0 & 238.3 & &  Sec.~\ref{ssec:vrate}, \ref{ssec:cacc} \\
MD3e1-HC10  & $0.3$  & 2.0 (10) & 238.3 (437.1) & 
 shock $\to$ photo. BC at $M_* = 10~M_\odot$ &  Sec.~\ref{ssec:cacc} \\
MD3e1-HC50  & $0.3$  & 2.0 (50) & 238.3 (826.8) &  
 shock $\to$ photo. BC at $M_* = 50~M_\odot$  &  Sec.~\ref{ssec:cacc} \\
MD1e1  & $0.1$  & 2.0 & 177.8 & fiducial case & Sec.~\ref{ssec:fid}, \ref{ssec:vrate} \\
MD6e2  & $0.06$  & 1.0 & 118.6 &  & Sec.~\ref{ssec:vrate} \\
MD3e2  & $0.03$  & 1.0 & 95.2  &  & Sec.~\ref{ssec:vrate} \\
MD6e3  & $0.006$  & 1.0 & 52.3  & also see \citet{OP03} & 
  Sec.~\ref{ssec:vrate}  \\
MD1e3  & $0.001$  & 0.05 & 12.5 & also see \citet{OP03} & 
  Sec.~\ref{ssec:fid}, \ref{ssec:vrate}  \\
\enddata
\tablecomments{Col.\ 2: mass accretion rate, Col.\ 3: initial stellar
mass, Col.\ 4: initial radius. 
For cases MD3e1-HC10 and MD3e1-HD50 the values when the boundary
condition is switched is given in parentheses.}
\end{deluxetable*}

\section{Numerical Modeling of Accreting Stars}
\label{sec:model}
\subsection{Method}
\label{sec:num}

We calculate stellar evolution with mass accretion using the 
numerical codes developed in our previous work 
\citep[see][for details]{OP03, HO09a, HYO10}.
The four stellar structure equations, i.e., 
equations of continuity, hydrostatic equilibrium, energy conservation, 
and energy transfer, including effects of mass 
accretion are solved assuming spherical symmetry. 
We focus on the early evolution until slightly after the 
ignition of hydrogen fusion in this paper.
To this end, the appropriate nuclear network for the
thermo-nuclear burning of deuterium, hydrogen,
and helium is considered.

The codes are designed to handle two 
different outer boundary conditions for stellar models: 
shock and photospheric boundaries. 
The shock boundary condition presupposes spherically symmetric
accretion onto a protostar, whereby the inflow directly
hits the stellar surface and forms a shock front \citep[e.g.,][]{SST80,HO09a}. 
We solve for the structure of both the stellar interior and outer
accretion flow. 
With this boundary condition the photosphere is located 
outside the stellar surface where the accretion flow is optically
thick to the stellar radiation.
The photospheric boundary condition, on the other hand, presupposes a 
limiting case of mass accretion via a circumstellar disk, whereby
accretion columns connecting the star and disk are geometrically
compact and most of the stellar surface radiates freely 
\citep[e.g.,][]{PS92,HYO10}. 
In this case we only solve the stellar interior structure without considering 
details of the accretion flow; the location of photosphere  
always coincides with the stellar surface.

The different outer boundary conditions correspond to the two
extremes of accretion flow geometries, or more specifically 
to different thermal efficiencies of 
mass accretion, which determine the specific entropy of 
accreting materials \citep[e.g.,][]{HOK11}. 
The accretion thermal efficiency controls the entropy content of 
the star, which determines the stellar structure. 
With the shock boundary condition, the accreting gas obtains a 
fraction of the entropy generated behind the shock front at the 
stellar surface. 
The resulting thermal efficiency is relatively high 
(``hot'' accretion). 

For the photospheric boundary condition, on the other hand, the accreting gas 
is assumed to have the same entropy as in the stellar atmosphere.
The underlying idea is that, when the accreting gas slowly
approaches the star via angular momentum transport in the disk, 
its entropy should be regulated to the atmospheric value.
This is a limiting case of thermally inefficient accretion
(``cold'' accretion).
In general, with even a small amount of angular momentum, 
mass accretion onto the protostar 
would be via a circumstellar disk, perhaps with geometrically
narrow accretion columns connecting
the disk with the star.
For extremely rapid mass accretion, however, the innermost part
of the disk becomes hot and entropy generated within the disk
is advected into the stellar interior \citep[e.g.,][]{Popham93}.
Thus, cold accretion as envisioned for the photospheric boundary 
condition is not appropriate for the case of rapid mass accretion
\citep[also see discussions in][]{HCK97,Smith11}.
We therefore expect that the shock boundary condition is a good 
approximation for the extremely high accretion rates
considered here and mostly focus on stellar 
evolution with the shock boundary condition.
We also consider a few cases with the photospheric boundary condition
for comparison to test potential effects of reducing the accretion
thermal efficiency (also see Sec.~\ref{sec:cases} below).

\subsection{Cases Considered}
\label{sec:cases}

The cases considered are summarized in Table 1.
In this paper, we only consider the evolution with constant
accretion rates for simplicity. The adopted accretion rates
range from $10^{-3}~\msunyr$ to $1~\msunyr$.
Stellar evolution with accretion rates less than 
$10^{-2}~\msunyr$ (cases MD1e3 and MD6e3) has been studied 
in detail in our previous work \citep[e.g.,][]{OP03,HO09a}.
As described in Section~\ref{sec:num} above, we calculate the
protostellar evolution assuming shock outer boundary conditions
for most of the cases. Cases MD3e1-HC10 and MD3e1-HC50 are
the only exceptions, whereby we switch to the photospheric boundary 
condition after
the stellar mass reaches $10~M_\odot$ and $50~M_\odot$,
respectively, at a constant accretion rate $0.3~\msunyr$.
The underlying idea for switching the boundary 
at some moment is that 
the specific angular momentum of the inflow 
and thus the circumstellar disk grows with time, 
which reduces entropy of the accreted matter.
The higher mass at switching point corresponds to 
the higher angular momentum of the parental core. 
As in our previous work, we start the calculations with initial 
stellar models constructed assuming that the stellar interior is 
in radiative equilibrium 
\citep[e.g.,][]{HO09a}.
We adopt a slightly higher initial stellar mass of $\sim 1 M_\odot$
for stability reasons.
The calculated initial stellar radii are also summarized in
Table \ref{tb:md}.

\section{Results}
\label{sec:results}

\subsection{Evolution in the Fiducial Case ($\mdot = 0.1~\msunyr$)}
\label{ssec:fid}

\begin{figure}
  \begin{center}
\epsscale{1.1}
\plotone{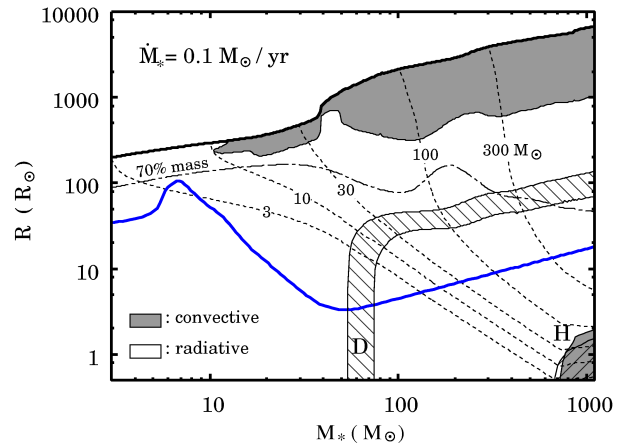}
\caption{Evolution of the stellar interior structure for 
the fiducial case, whereby the stellar mass increases at a rate
of $0.1~\msunyr$ (case MD1e1). 
The thick solid line depicts the stellar surface, which is the
position of the accretion shock front.
The dotted lines show the radial positions of the mass 
coordinates of $M = 3$, 10, 30, 100, and $300~M_\odot$.
The dot-solid line indicates the radial position within which
70~\% of the stellar mass is enclosed. 
The gray-shaded areas represent the convective layers.
The hatched areas indicate layers of active nuclear burning,
where the energy production rate exceeds 10\% of the steady rate
$0.1 L_{\rm D,st}/M_{\ast}$ for deuterium burning (see eq. \ref{eq:LDst}), 
and $0.1 L_{\ast}/M_{\ast}$ for hydrogen burning. 
The blue solid line shows the evolution of the radius
of a star accreting material at $10^{-3}~\msunyr$ for comparison.
}
\label{fig:interior_MD1e1}
  \end{center}
\end{figure}
\begin{figure}
  \begin{center}
\epsscale{1.0}
\plotone{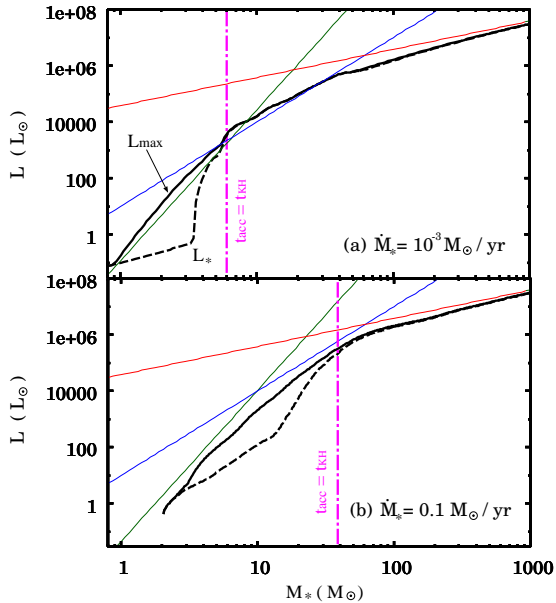}
\caption{Evolution of the stellar surface luminosity $L_*$ (dashed line) 
and maximum luminosity within the star $L_{\rm max}$ (solid line) 
for the cases with $10^{-3}~\msunyr$ (MD1e3; {\it upper panel}) and 
$0.1~\msunyr$ (fiducial case MD1e1; {\it lower panel}).
The mass-luminosity relations given by equations
(\ref{eq:lmax_kramers}), (\ref{eq:lmax_esc}) and (\ref{eq:ledd})
are shown with the thin green, blue, and red lines, respectively.
In each panel the vertical dot-dashed line (magenta) indicates the epoch when
the accretion time is equal to the KH time.
}
\label{fig:lmax}
  \end{center}
\end{figure}
\begin{figure}
  \begin{center}
\epsscale{1.0}
\plotone{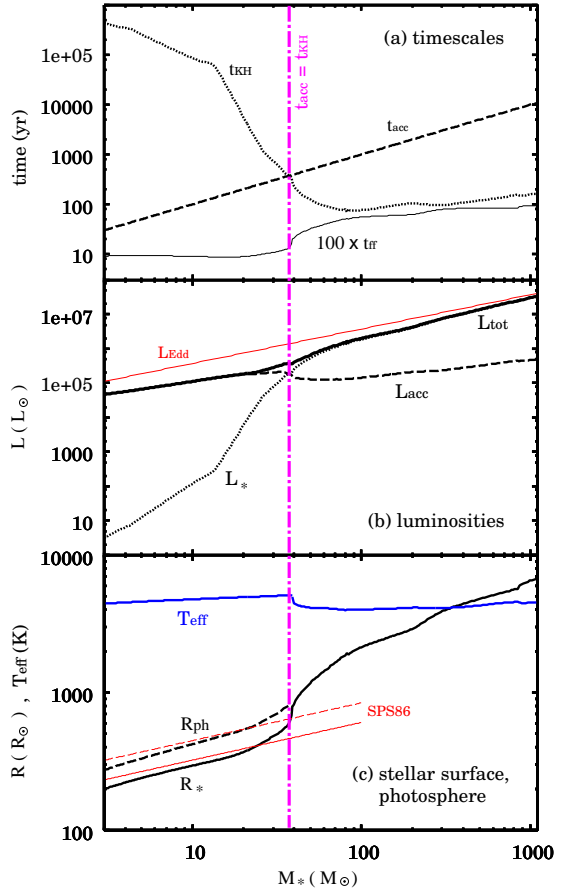}
\caption{
Evolution of several physical quantities 
for the fiducial case with $0.1~\msunyr$ (MD1e1).
{\it Top panel:} 
Comparison between the accretion timescale $t_{\rm acc}$ (dashed line) 
and KH timescale $t_{\rm KH}$ (dotted line).
The vertical magenta dot-dashed line indicates the epoch
when $t_{\rm KH}$ is equal to $t_{\rm acc}$.
The thin solid line represents 100 times of the stellar 
free-fall timescale $t_{ff} \equiv \sqrt{3 \pi / 32 G \bar{\rho}}$, 
where $\bar{\rho}$ is the average mass density of the star.
The fact that $t_{ff}$ is much shorter than $t_{\rm KH}$ and 
$t_{\rm acc}$ verifies the hydrostatic balance assumption 
implicit in the stellar structure equations.
{\it Middle panel:} 
Evolution of the accretion luminosity $L_{\rm acc}$ (dashed line), 
stellar luminosity $L_*$ (dotted line), 
and total luminosity $L_{\rm tot} \equiv L_{\rm acc} + L_*$ (solid line).
The red line indicates the Eddington luminosity at each stellar mass.
{\it Bottom panel:} 
Evolution of the radial positions of the photosphere 
$R_{\rm ph} (R_\odot)$ (dashed line)
and stellar surface $R_* (R_\odot)$ (black solid line).
The red solid and dashed lines denote the analytic formulae
for these radii by \citet{SPS86} (eqs \ref{eq:rst_sps} and \ref{eq:rph_sps}). 
The evolution of the stellar effective temperature $T_{\rm eff} (K)$
is also shown with the blue line.
}
\label{fig:lumtsc_MD1e1}
  \end{center}
\end{figure}

We first consider the fiducial case (MD1e1), whereby the stellar mass 
increases with the constant accretion rate of $\mdot = 0.1~\msunyr$. 
The calculated evolution of the stellar interior structure is 
presented in Figure \ref{fig:interior_MD1e1}. 
We see that the stellar radius is very large and 
increases monotonically with the stellar mass. 
The stellar radius exceeds $10^3~R_\odot$ when the stellar mass
is $M_* \simeq 45~M_\odot$ and reaches $\simeq 6500~R_\odot$ at
$M_* \simeq 10^3~M_\odot$.
This evolution differs qualitatively from that calculated assuming a
lower accretion rate $10^{-3}~\msunyr$ as depicted 
in Figure \ref{fig:interior_MD1e1} by the blue line 
\citep[see also e.g.,][]{OP03}. 
At the lower accretion rate the stellar radius initially increases with
stellar mass but begins to decrease at $M_* \gtrsim 6~M_\odot$.

A key quantity for understanding the contrast between the two cases
is the balance between
the two characteristic timescales: the accretion timescale
\begin{equation}
t_{\rm acc} \equiv \frac{M_*}{\dot{M}_*}
\end{equation}
and Kelvin-Helmholtz (KH) timescale
\begin{equation}
t_{\rm KH} \equiv \frac{G M_{\ast}^2}{R_{\ast} L_{\ast}} ,
\end{equation}
where $R_*$ and $L_*$ are the stellar radius and luminosity, and
$G$ is the gravitational constant
\citep[e.g.,][]{SPS86,OP03,HO09a}. 
In the early stage, during which the stellar radius increases with mass, 
the timescale balance is $t_{\rm acc} \ll t_{\rm KH}$ and radiative 
loss of the stellar energy is negligible ({\it adiabatic accretion stage}).
However, the radiative energy loss becomes more efficient
as the stellar mass increases. 
This is because opacity in the stellar interior, which is due to
the free-free absorption following Kramers' law 
$\kappa \propto \rho T^{-3.5}$, decreases
and the stellar luminosity $L_*$ increases with the stellar interior 
temperature (and thus with its mass $M_*$).
The upper panel of Figure~\ref{fig:lmax} indeed shows that 
the maximum luminosity within the star $L_{\rm max}$ 
increases as a power-law function of $M_*$. 

This increase of $L_*$ is consistent with the analytic
scaling relation for radiative stars with Kramers' opacity,
$L \propto M_*^{11/2} R_*^{-1/2}$ \citep[e.g.,][]{HHS62}. 
Our numerical results are well fitted by the analytic relation
\begin{equation}
L_{\rm max} \simeq 0.6~L_\odot 
\left( \frac{M_*}{M_\odot} \right)^{11/2}
\left( \frac{R_*}{R_\odot}  \right)^{-1/2} .
\label{eq:lmax_kramers}
\end{equation}
The increase of $L_*$ causes an inversion of the timescale balance 
to $t_{\rm acc} > t_{\rm KH}$ at low accretion rates. The star contracts 
by losing its energy ({\it KH contraction stage}), which is
seen for $M_* \gtrsim 6~M_\odot$.
The opacity in the stellar interior has fallen down to the
constant value of electron scattering.
Figure~\ref{fig:lmax} shows that, in this stage, luminosity takes 
its maximum value at the stellar surface and increases as 
$L_{\ast}=L_{\rm max} \propto M_*^3$, which is valid for the  
constant opacity cases \citep[e.g.,][]{HHS62}. The relation
\begin{equation}
L_{\rm max} \simeq 10~L_\odot \left( \frac{M_*}{M_\odot} \right)^3
\label{eq:lmax_esc}
\end{equation}
roughly agrees with our results.
Temperature at the stellar center increases during the KH contraction stage.
Hydrogen burning finally begins and the stellar radius begins to
increase following the mass-radius relation of ZAMS stars
for $M_* \gtrsim 50~M_\odot$.
Figure~\ref{fig:lmax} shows that the stellar luminosity gradually 
approaches to the Eddington limit
\begin{equation}
L_{\rm Edd} (M_*) \simeq 3.8 \times 10^6~L_\odot 
\left( \frac{M_*}{100~M_\odot} \right) .
\label{eq:ledd}
\end{equation}

\setcounter{footnote}{0}

By contrast, there is no contraction stage 
for the case with a much higher accretion rate
$\mdot = 0.1~\msunyr$ (MD1e1).
Nevertheless, the evolution of the timescales still 
follows the picture described above (Fig.~\ref{fig:lumtsc_MD1e1} a).
We see that the timescale balance changes from $t_{\rm acc} < t_{\rm KH}$
to $t_{\rm acc} > t_{\rm KH}$ at $M_* \simeq 40~M_\odot$.
The protostar is in the adiabatic accretion stage for 
$M_* \lesssim 40~M_\odot$.
The accretion luminosity $L_{\rm acc} \equiv G M_* \mdot / R_*$ at this stage
is much higher than the stellar luminosity $L_*$, 
since the luminosity ratio $L_{\rm acc}/L_*$ is equal to 
the timescale ratio $t_{\rm KH}/t_{\rm acc}$ by definition.
\citet{SPS86} derived the approximate analytic formulae describing 
radial positions of the stellar surface $R_*$ and 
photosphere $R_{\rm ph}$ (located within the accretion flow)
during the adiabatic stage:
\begin{equation}
R_* \simeq 26~R_\odot \left( \frac{M_*}{M_\odot}  \right)^{0.27}
\left( \frac{\mdot}{10^{-3}~\msunyr}  \right)^{0.41} ,
\label{eq:rst_sps}
\end{equation}
\begin{equation}
R_{\rm ph} \simeq 1.4 R_* .
\label{eq:rph_sps}
\end{equation}

The bottom panel of Figure~\ref{fig:lumtsc_MD1e1} shows that these 
formulae still agree with our numerical results with 
$0.1~\msunyr$.
Equation (\ref{eq:rst_sps}) shows that the stellar radius is larger
for the higher accretion rate at the same mass. 
The larger radius is due to the higher specific entropy of accreting material
and the resulting higher entropy content of the star \citep[e.g.,][]{HO09a}.
Comparing two stars of the same mass, the one with a larger radius 
has a lower interior temperature, which then implies a higher opacity
due to the strong $T$-dependence of Kramers' law 
$\kappa \propto \rho T^{-3.5}$.
For this reason the adiabatic accretion stage is prolonged up to
higher stellar masses for higher accretion rates.
The stellar luminosity $L_*$ thus increases with stellar mass.
Figure~\ref{fig:lmax} (b) shows that the evolution of the 
stellar maximum luminosity $L_{\rm max}$ still obeys equations 
(\ref{eq:lmax_kramers}) - (\ref{eq:ledd}).
Unlike in the case for $10^{-3}~\msunyr$, however, 
it is only after $L_{\rm max}$ approaches 
the relation of $L_{\rm max} \propto M_*^3$ (eq. \ref{eq:lmax_esc})
that $t_{\rm KH}$ becomes equal to $t_{\rm acc}$.
The rapid heat input via mass accretion prevents the star from
losing internal energy until the stellar luminosity becomes sufficiently high.

\begin{figure}
  \begin{center}
\epsscale{1.0}
\plotone{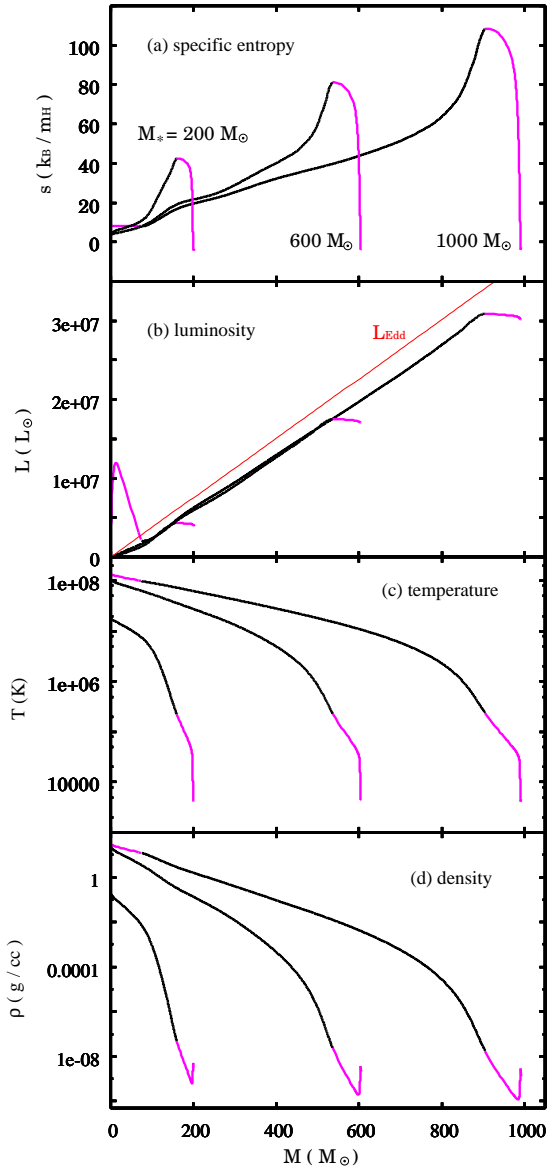}
\caption{Radial profiles of the specific entropy ({\it panel a}), 
luminosity ({\it b}), temperature ({\it c}) and gas mass density
({\it d}) in the stellar interior 
for the fiducial case with $0.1~\msunyr$ (MD1e1).
The profiles when the stellar mass is $200~M_\odot$, $600~M_\odot$,
and $10^3~M_\odot$ are shown in each panel.
The magenta parts indicate the layers are convective.
The thin red line in panel (b) represents the Eddington-limit 
luminosity as a function of the mass coordinate $M$.
}
\label{fig:struct_MD1e1}
  \end{center}
\end{figure}

The fact that $t_{\rm KH}$ is shorter than $t_{\rm acc}$ 
for $M_* \gtrsim 40~M_\odot$ indicates that most of 
the stellar interior is contracting, as shown by 
the trajectories of the mass coordinates (dashed
lines in Figure \ref{fig:interior_MD1e1}). 
The figure also shows that the bloated surface layer occupies only
a small fraction of the total stellar mass. 
When the stellar mass is $\simeq 300~M_\odot$, 
for example, the layer which has 30~\% of the total mass measured
from the surface covers more than 98~\% of the radial extent. 
The star has a radiative core surrounded by an outer
convective layer. 
Although the convective layer covers a large fraction of 
the stellar radius, even the central radiative core is much larger 
than a ZAMS star with the same mass
(compare with the blue curve for $M_* \gtrsim 40~M_\odot$ 
in Fig.~\ref{fig:interior_MD1e1}).
Figure \ref{fig:struct_MD1e1} shows the radial distributions 
of physical quantities, i.e., specific entropy, luminosity, temperature, 
and density, in the stellar interior.
We see that the specific entropy is at its maximum value near the
boundary between the radiative core and convective layer.
The stellar entropy distribution is controlled
by the energy equation,
\begin{equation}
T \left( \frac{\partial s}{\partial t}  \right)_M = 
\epsilon - 
\left( \frac{\partial L}{\partial M}  \right)_t,
\label{eq:energy}
\end{equation}
where $s$ is the specific entropy and 
$\epsilon$ is the energy production rate by nuclear fusion.
For $M_* \gtrsim 40~M_\odot$ most of the stellar luminosity comes 
from the release of gravitational energy. In fact, 
as seen in Figure~\ref{fig:struct_MD1e1} (b),
$(\partial L/\partial M)_t > 0$ in the radiative core, 
which means that the internal energy of the gas is decreasing. 
The local luminosity in the radiative core is close to 
the Eddington value given by equation (\ref{eq:ledd}), using 
the mass coordinate $M$ rather than stellar mass $M_*$ 
(Fig.~\ref{fig:struct_MD1e1} b). Note that the opacity in the 
radiative core is only slightly higher than that expected from 
electron-scattering alone.

In the outer parts of the star, where temperature and density are lower,
however, opacity is higher than in the core because of
bound-free absorption of H, He atoms and the H$^-$ ion.
Energy transport via radiation is inefficient there, and
a part of the energy coming from the core is carried 
outward via convection.
Figure~\ref{fig:struct_MD1e1} shows that the surface convective layer
lies in the temperature and density range 
where $T \lesssim 10^5$~K and $\rho \lesssim 10^{-8}~{\rm cm}^{-3}$,
which is almost independent of the stellar mass.
We also see that the specific entropy is not constant over
the convective layer, decreasing toward the stellar surface 
(a so-called {\it super-adiabatic} layer).
This is because convective heat transport 
is inefficient near the stellar surface.
A part of the outflowing energy is absorbed there,
as indicated by the fact that the surface layer has a
negative luminosity gradient $(\partial L/\partial M)_t < 0$.
This explains the high specific entropy in the outer convective layer.

The outermost part of the star has a density inversion,
i.e. the density increases toward the stellar surface.
Here, opacity assumes very high values because of
H$^-$ absorption. Radiation pressure is so strong that 
the hydrostatic balance is not achieved only with gravity;
the additional inward force by the negative gas pressure gradient 
helps maintain the hydrostatic structure.  Note, however, that this 
density inversion could be unstable in realistic
multi-dimensions \citep[see e.g.,][]{Begelman08}.

Although deuterium burning is ignited when the stellar mass is 
$\simeq 50~M_\odot$, its influence on the subsequent
evolution is negligible (Fig.~\ref{fig:interior_MD1e1}).
The total energy production rate by deuterium burning is approximately
\begin{eqnarray}
L_{\rm D,st} &\equiv& \mdot \delta_{\rm D} \\
&=& 1.5 \times 10^5~L_\odot \left( \frac{\mdot}{0.1~\msunyr} \right)
\left( \frac{[{\rm D/H}]}{2.5 \times 10^{-5}} \right) 
\nonumber,
\label{eq:LDst}
\end{eqnarray}
where $\delta_{\rm D}$ is the energy available from deuterium
burning per unit gas mass. 
Since this is much lower than the Eddington luminosity $L_{\rm Edd}$ 
in the mass range considered, energy production by deuterium burning
contributes only slightly to the luminosity in the stellar interior.
\citet{HO09a} showed that deuterium burning influences the stellar
evolution only when the accretion rate is low 
$\mdot \lesssim 10^{-4}~\msunyr$.

Figure~\ref{fig:struct_MD1e1} (c) shows that temperature in the
stellar interior increases with total mass.
The central temperature reaches $10^8$~K when the stellar
mass is $\simeq 600~M_\odot$.
Soon after that, hydrogen burning begins and a central 
convective core develops (Fig.~\ref{fig:interior_MD1e1}).
This convective core can also be seen in the radial profiles for the  
$10^3~M_\odot$ model in Figure~\ref{fig:struct_MD1e1} (indicated by magenta).
The luminosity profile tells that most of the energy
produced by hydrogen burning is absorbed within the convective core.
The star still shines largely by releasing its gravitational energy even 
after hydrogen ignition. 


When the stellar radius is sufficiently large, the accretion flow 
reaches the stellar surface before becoming opaque 
to the outgoing stellar light. 
In fact, soon after the end of the adiabatic accretion stage, 
the accreting envelope remains optically thin throughout
(Fig.~\ref{fig:lumtsc_MD1e1} c). 
We also see that the stellar effective temperature 
is almost constant at $T_{\rm eff} \simeq 5000$~K during this period.
In general, the stellar effective temperature never assumes
a lower value due to the strong temperature-dependence of
H$^-$ absorption opacity \citep[e.g.,][]{Hayashi61}.
Stars that have a compact core and bloated envelope
(e.g., red giants) commonly have an almost constant effective 
temperature, regardless of their stellar masses.

\subsection{Cases with Different Accretion Rates}
\label{ssec:vrate}

\begin{figure}
  \begin{center}
\epsscale{1.1}
\plotone{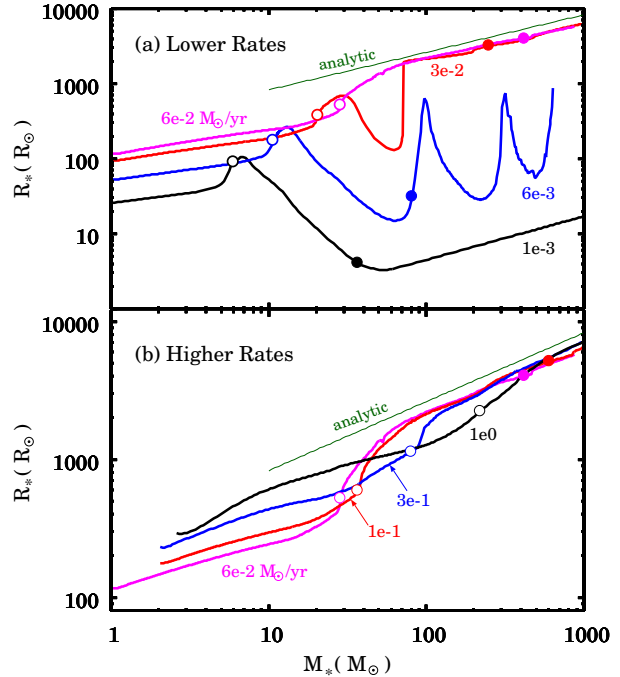}
\caption{Evolution of the protostellar radius for various accretion rates.
{\it Upper panel:} the different curves represent the cases with 
$\mdot = 10^{-3}~\msunyr$ (case MD1e3, black), $6 \times 10^{-3}~\msunyr$ 
(MD6e3, blue), $3 \times 10^{-2}~\msunyr$ (MD3e2, red), and
$6 \times 10^{-2}~\msunyr$ (MD6e2, magenta).
The open and filled circles on each curve denote the epoch
when $t_{\rm KH} = t_{\rm acc}$ and when the
central hydrogen burning begins, respectively.
{\it Lower panel:} same as the upper panel but for higher accretion
rates of $6 \times 10^{-2}~\msunyr$ (MD6e2, magenta), 
$0.1~\msunyr$ (MD1e1, red), $0.3~\msunyr$ (MD3e1, blue), 
and $1~\msunyr$ (MD1e0, black). 
The case MDe2 is illustrated in both panels as a reference. 
For the cases with $0.3~\msunyr$ and $1~\msunyr$ (MD3e1 and MD1e0)
hydrogen fusion has not ignited
by the time the stellar mass reaches $10^3~M_\odot$.
In the both panels the thin green line represents the mass-radius
relation given by equation (\ref{eq:rst_analytic}).
}
\label{fig:m_r}
  \end{center}
\end{figure}
\begin{figure}
\epsscale{1.0}
\plotone{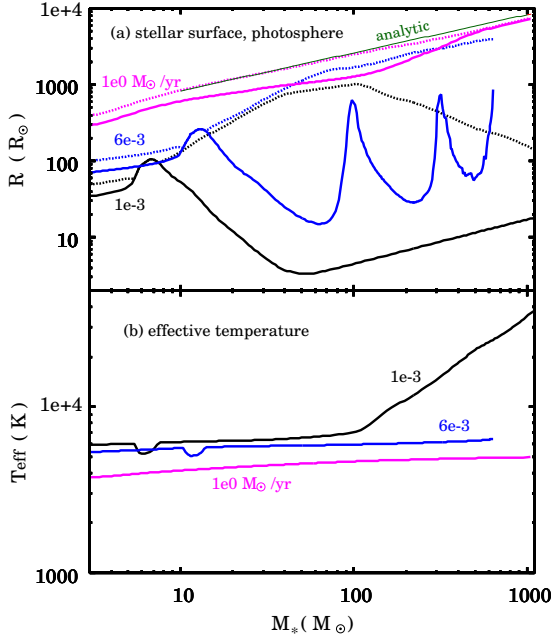}
\caption{
{\it Upper panel:} Evolution of radial positions of the stellar
surface (solid) and photosphere (dotted). The black, blue,
and magenta curves denote cases with $\mdot = 10^{-3}~\msunyr$
(case MD1e3), $6 \times 10^{-3}~\msunyr$ (MD6e3), and
$1~\msunyr$ (MD1e0), respectively.
The mass-radius relation given by equation (\ref{eq:rst_analytic}) 
is plotted with the thin green line.
{\it Lower panel:} The stellar effective temperature for the same cases 
as in the upper panel.
}
\label{fig:prec_teff}
\end{figure}
\begin{figure}
\epsscale{1.0}
\plotone{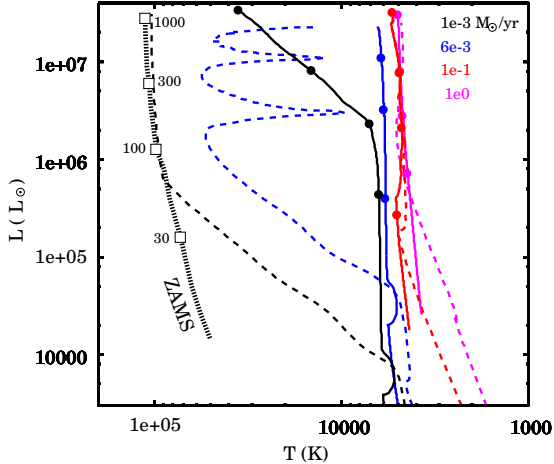}
\caption{Evolutionary tracks in the HR diagram. The different colors denote
the evolution for different accretion rates, $10^{-3}~\msunyr$ 
(case MD1e3, black), $6 \times 10^{-3}~\msunyr$ (MD6e3, blue),
$0.1~\msunyr$ (MD1e1, red), $1~\msunyr$ (MD1e0, magenta). 
For each case the values at the stellar photosphere 
($L_{\rm tot}$ and $T_{\rm eff}$) and at the stellar
surface ($L_*$ and $T_{\rm eff,*} \equiv L_* / 4 \pi \sigma R_*^2$) 
are plotted with solid and dashed lines, respectively.
The thick dashed line represents the loci of non-accreting 
ZAMS stars taken from \citet{Marigo01}
($M_* \leq 100~M_\odot$) and \citet{Bromm01} ($M_* \geq 100~M_\odot$). 
The filled circles and open squares on the lines mark the positions
for $M_* = 30~M_\odot$, $100~M_\odot$, $300~M_\odot$, and
$10^3~M_\odot$ in ascending order.}
\label{fig:HR}
\end{figure}
\begin{figure}
  \begin{center}
\epsscale{1.0}
\plotone{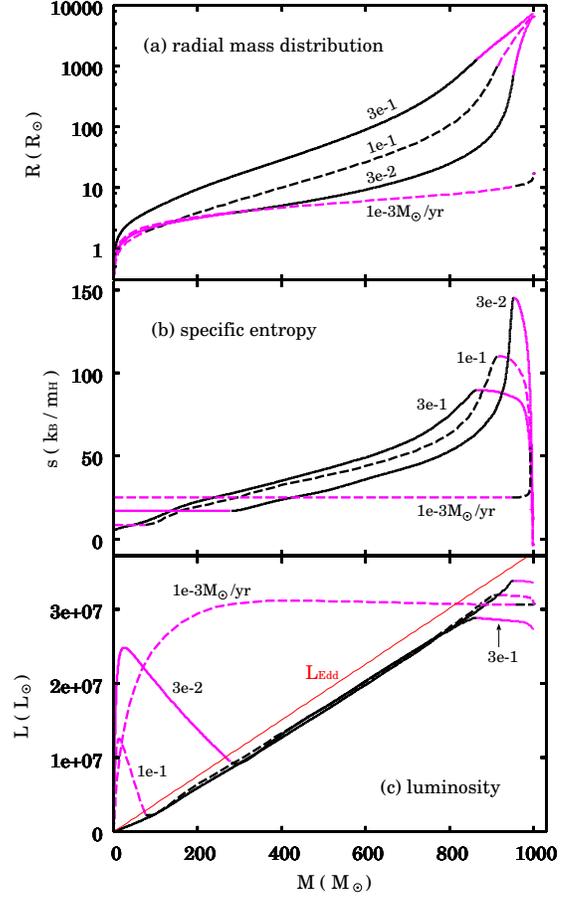}
\caption{Comparisons of the interior structure of $10^3~M_\odot$
stars produced by different accretion rates.
Radial profiles of mass ({\it panel a}), specific entropy ({\it b}),
and luminosity ({\it c}) are shown. 
In each panel the solid and dashed curves represent the cases
for $0.3~\msunyr$ (MD3e1), $0.1~\msunyr$ (MD1e1),  
$3 \times 10^{-2}~\msunyr$ (MD3e2), and $10^{-3}~\msunyr$ (MD1e3).
The magenta lines mark convective layers in the stellar interior.
The thin red line in panel (c) represents the Eddington luminosity
as a function of mass $M$.
}
\label{fig:struct_1000m}
  \end{center}
\end{figure}

We now investigate how stellar evolution changes with 
the accretion rate. 
Figure~\ref{fig:m_r} shows the evolution of the 
stellar radius for several cases, including 
$\mdot = 10^{-3}~\msunyr$ (case MD1e3) and 
$0.1~\msunyr$ (case MD1e1) explained in Section~\ref{ssec:fid}.
We see that, with accretion rates higher than 
$6 \times 10^{-2}~\msunyr$, the evolution becomes similar to
that of the fiducial case for $0.1 ~\msunyr$ (MD1e1); 
the stellar radius monotonically increases with mass.
The protostars undergo adiabatic accretion in the early stage.
As equation (\ref{eq:rst_sps}) shows, the stellar radius is larger
for higher accretion rates at a given stellar mass, say, at
$M_* = 10~M_\odot$.

Adiabatic accretion occurs up to a higher stellar mass
when the accretion rate is higher.
We derive an analytic expression describing this dependence in the following.
As long as the KH timescale $t_{\rm KH}$ is longer than
the accretion timescale $t_{\rm acc}$, we have adiabatic accretion. 
\citep[e.g.,][]{HO09a}. 
Figure~\ref{fig:lmax} shows that $L_{\rm max}$, whose evolution 
is well described by simple analytic expressions 
(eqs. \ref{eq:lmax_kramers} and \ref{eq:lmax_esc}), converges to 
$L_*$ at the end of the adiabatic accretion stage. 
Thus, the stellar mass when the adiabatic accretion terminates 
$M_{\rm *, teq}$ can be estimated with the relation
$t_{\rm acc} \simeq t_{\rm KH} = G M_*^2 / R_* L_{\rm max}$.
Eliminating $L_{\rm max}$ and $R_*$ with equations 
(\ref{eq:lmax_esc}) and (\ref{eq:rst_sps}), we obtain
\begin{equation}
M_{\rm *, teq} \simeq 14.9~M_\odot
\left( \frac{\mdot}{10^{-2}~\msunyr}  \right)^{0.26} .
\label{eq:mteq}
\end{equation}
We have confirmed that the epoch of the timescale equality in 
our numerical calculation is well described by this equation.


Even after $t_{\rm KH}$ becomes shorter than $t_{\rm acc}$, 
the stellar radius continues to increase for accretion rates 
$\ga 10^{-2}~\msunyr$. 
The variations of radii among cases with 
different accretion rates gradually disappear.
The stellar radii finally converge to a unique mass-radius
relation with $R_* \propto M_*^{1/2}$ in all the cases.
We can derive the approximate mass-radius relation 
from the following simple argument. 
First, the stellar luminosity is generally written as
\begin{equation}
L_* = 4 \pi R_*^2 \sigma T_{\rm eff}^4 ,
\label{eq:lst_def}
\end{equation}
where $\sigma$ is the Stefan-Boltzman constant. 
As Figure~\ref{fig:lumtsc_MD1e1} indicates, 
the stellar luminosity 
approaches the Eddington value $L_{\rm Edd}(M_*)$ 
for $M_* \ga 100~M_\odot$. 
As explained in Section~\ref{ssec:fid} the stellar
effective temperature stays at the constant value 
$T_{\rm eff} \simeq 5000$~K after the inversion of the timescales
(also see Figs.~\ref{fig:prec_teff} and \ref{fig:HR}).
Substituting these relations into equation (\ref{eq:lst_def}), 
we obtain
\begin{equation}
R_* \simeq 2.6 \times 10^3~R_\odot 
\left( \frac{M_*}{100~M_\odot} \right)^{1/2} .
\label{eq:rst_analytic}
\end{equation}
Figure~\ref{fig:m_r} shows that our numerical results 
approximately follow this relation.

\citet{Begelman10} also considered stellar evolution with
very rapid mass accretion using simple analytic arguments.
His model predicts that the stellar radius is 
proportional to the mass accretion rate (his eq. 24), 
which does not agree with our numerical results.
\citet{Begelman10}, however, does not take into account
the detailed structure of the outermost layer of the star, 
where H$^-$ opacity is important.
The fact that the strong $T$-dependence of H$^-$ opacity 
keeps the stellar effective temperature almost constant is
essential for our results. 


Cases MD3e2 and MD6e3 model stellar evolution at 
intermediate accretion rates 
($3\times 10^{-2}$ and $6\times 10^{-3}~\msunyr$, respectively)
and exhibit a different behavior 
than the higher accretion-rate cases described above. 
For an accretion rate $3\times 10^{-2}~\msunyr$ (MD3e2) 
the protostar initially contracts after the adiabatic accretion stage. 
At $M_* \simeq 70~M_\odot$, however, the stellar radius sharply
increases  and ultimately converges to
the mass-radius relation given by equation (\ref{eq:rst_analytic}).
The case with $6\times 10^{-3}~\msunyr$ exhibits an oscillatory behavior 
of the stellar radius for $M_* \gtrsim 70~M_\odot$.
In this case the accreting envelope remains
optically thick after the onset of KH contraction
(Figs.~\ref{fig:prec_teff} and \ref{fig:HR}). Its photospheric radius 
still follows the mass-radius relation $R_{\rm ph} \propto M_*^{1/2}$.
The effective temperature also assumes the constant value
$T_{\rm eff} \simeq 6000~K$.
These evolutionary features for 
$\mdot < 10^{-2}~\msunyr$ have 
also been found in previous studies \citep[e.g.,][]{OP01,OP03}. 

We have seen that the stellar radius at $M_* \simeq 10^3~M_\odot$ is
almost independent of accretion rate 
as long as $\mdot \gtrsim 3 \times 10^{-2}~\msunyr$.
However, the stellar interior structure at this moment is not
identical among these cases (Fig.~\ref{fig:struct_1000m}).
Although each of these stars has a radiative core and 
a convective envelope, the mass is more strongly centrally 
concentrated for the lower accretion rates;
the less massive envelopes have a higher entropy and
inflate even more to achieve the same stellar radius.

\begin{figure}
  \begin{center}
\epsscale{1.1}
\plotone{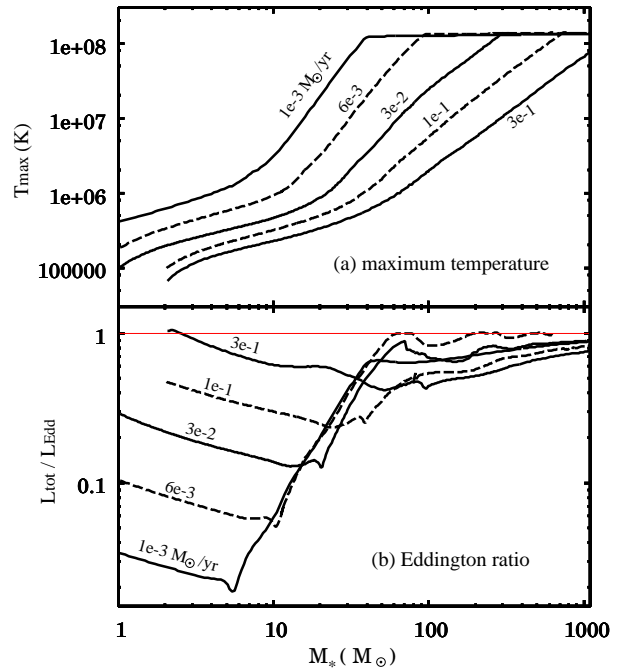}
\caption{Evolution of the maximum temperature in the stellar interior
({\it upper panel}) and Eddington ratio $L_{\rm tot}/\ledd$
({\it lower panel}) with increasing stellar mass.
The solid and dashed curves alternately represent the cases
with different accretion rates, $10^{-3}~\msunyr$ (case MD1e3),
$6 \times 10^{-3}~\msunyr$ (MD6e3), $3 \times 10^{-2}~\msunyr$
(MD3e2), $0.1~\msunyr$ (MD1e1), and $0.3~\msunyr$ (MD3e1).
}
\label{fig:tmax_fedd}
  \end{center}
\end{figure}

The evolution of the stellar maximum temperature is 
helpful for understanding the variation of the stellar
interior structure (Fig.~\ref{fig:tmax_fedd} a) with accretion rate.
As equation (\ref{eq:mteq}) shows, the central part of the star 
begins to contract and release gravitational energy 
at a lower stellar mass for the lower accretion rates. 
The central temperature quickly increases 
with stellar mass once the KH time becomes
shorter than the accretion time.
The maximum temperature $T_{\rm max}$ reaches $10^8$~K at a lower 
stellar mass for the lower accretion rate. After that $T_{\rm max}$ 
assumes an almost constant value due to the strong $T$-dependence 
of the energy production rate of hydrogen burning. 
For the case with $3\times 10^{-2}~\msunyr$ (MD3e2) hydrogen burning
begins at $M_* \simeq 200~M_\odot$;
the resulting central convective core is seen in the profiles
in Figure \ref{fig:struct_1000m}. This feature is not seen for 
the case with $0.3~\msunyr$, because hydrogen has not yet ignited 
by the time $M_* = 10^3~M_\odot$ for $\mdot \gtrsim 0.3~\msunyr$.

Figure~\ref{fig:m_r} shows that for 
$\mdot \gtrsim 6 \times 10^{-3}~\msunyr$ the protostar cannot 
reach the ZAMS stage by KH contraction. 
\cite{OP03} pointed out that this is because the total
luminosity $L_{\rm tot} \equiv L_* + L_{\rm acc}$
becomes close to the Eddington limit during 
the contraction to the ZAMS. 
For the cases with $6 \times 10^{-3}$ and 
$3 \times 10^{-2}~\msunyr$ (cases MD6e3 and MD3e2),
for example, the abrupt expansion 
terminates the KH contraction when the total luminosity is
nearly at the Eddington limit (Fig.~\ref{fig:tmax_fedd} b).
\cite{OP03} analytically derived the maximum accretion rate
$\simeq 4 \times 10^{-3}~\msunyr$ 
with which the protostar can reach the ZAMS following 
KH contraction. Figure \ref{fig:m_r} indicates that 
there is another critical accretion rate 
$\simeq 6 \times 10^{-2}~\msunyr$, above 
which the stellar evolution changes qualitatively;
the KH contraction stage disappears entirely at higher rates.
This critical rate can also be derived from a similar argument
as the one above.

Note that the increase of stellar mass during 
the KH contraction stage is smaller for the case with 
$3 \times 10^{-2}~\msunyr$ than with $6 \times 10^{-3}~\msunyr$. 
Extending this fact for our critical case, 
the total luminosity would nearly reach the Eddington limit 
just at the end of the adiabatic accretion stage, i.e., when
$t_{\rm KH} \simeq t_{\rm acc}$. 
Since the opacity in the surface layer is higher than 
from Thomson scattering during this epoch, 
having the total luminosity only slightly lower than the Eddington value 
causes the star to expand. 
Thus, the condition for the critical case is
\begin{equation}
2 L_{\rm max} \simeq C_{\rm Edd} L_{\rm Edd},
\label{eq:ltot_ledd}
\end{equation}
where $C_{\rm Edd}$ is a factor less than the unity 
and we have used
the fact that the total luminosity is written as 
$L_{\rm tot} \simeq 2 L_{\rm max}$ when $t_{\rm KH} \simeq t_{\rm acc}$. 
Using $C_{\rm Edd} = 0.25$ as a fiducial value 
(Fig.~\ref{fig:tmax_fedd} b) and equation (\ref{eq:lmax_esc}) 
for $L_{\rm max}$, the stellar mass which satisfies 
the condition (\ref{eq:ltot_ledd}) is
\begin{equation}
M_{\rm *,Edd,teq} \simeq 21.7~M_\odot 
\left( \frac{C_{\rm Edd}}{0.25} \right)^{0.5} .
\label{eq:medd_teq}
\end{equation}
On the other hand, equation (\ref{eq:mteq}) also gives 
the stellar mass when 
$t_{\rm KH} \simeq t_{\rm acc}$ for a given accretion rate. 
Equating $M_{\rm *, teq}$ and $M_{\rm *.Edd,teq}$ with equations
(\ref{eq:mteq}) and (\ref{eq:medd_teq}), we obtain the critical mass
accretion rate
\begin{equation}
\dot{M}_{\rm cr} \simeq 4.7 \times 10^{-2}~\msunyr
\left( \frac{C_{\rm Edd}}{0.25}  \right)^{1.9} ,
\end{equation}
which agrees with our numerical results.

%

\subsection{Effects of Lower-Entropy Accretion}
\label{ssec:cacc}

\begin{figure}
\epsscale{1.0}
\plotone{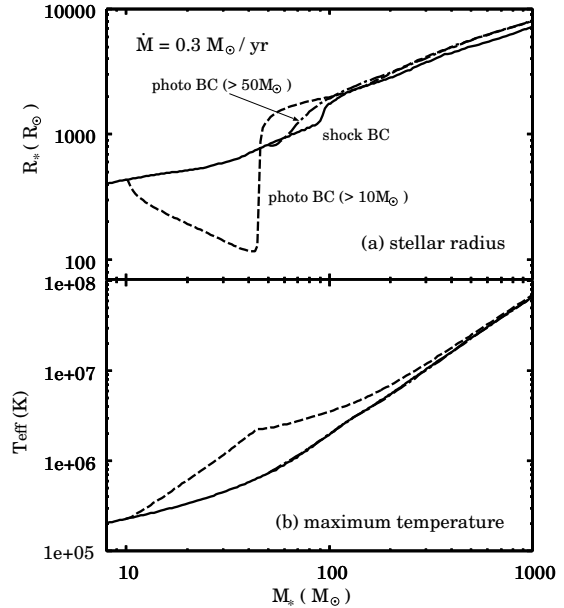}
\caption{Effect of reducing the thermal efficiency of mass accretion 
({\it upper panel:} stellar radius, {\it lower panel:} maximum
temperature within the star).
The same accretion rate of $0.3~\msunyr$ is adopted for
all three cases presented. 
The solid line represents the evolution with the 
shock boundary condition, i.e, thermally efficient or ``hot'' 
accretion, throughout (MD3e1). 
The dashed and dot-dashed lines show the evolution in cases 
MD3e1-HC10 and MD3e1-HC50, where the photospheric boundary condition 
(i.e., thermally inefficient or ``cold'' accretion) is adopted 
after the stellar mass exceeds $10~M_\odot$ and $50~M_\odot$, respectively.
In the lower panel the dot-dashed line is indistinguishable from
the solid line.
}
\label{fig:mr_tmax_pb}
\end{figure}

We have used the shock outer boundary condition for the 
stellar models presented and discussed above. 
As discussed in Section \ref{sec:num}, the shock boundary condition,
which implies that the accreting gas joins the star with relatively
high entropy, would be valid for cases with the very rapid 
mass accretion considered in this paper.
If the accred gas had lower entropy, however,  
the stellar radius would be reduced because of the resulting lower
entropy thoughout the stellar interior.
Here, we examine potential effects of the colder mass accretion 
by adopting the photospheric boundary conditions 
\citep[e.g.,][]{HYO10,HOK11}. 
Figure~\ref{fig:mr_tmax_pb} (a) shows the evolution of the stellar 
radius for three cases with $0.3~\msunyr$, whereby the shock 
boundary condition is used throughout in one case (MD3e1), 
whereas the outer boundary condition is changed to 
the photospheric one for $> 10~M_\odot$ (MD3e1-HCm10) 
and $>50~M_\odot$ (MD3e1-HCm50), respectively. 
The stars are still in the adiabatic accretion stage when the 
boundary condition is changed in both cases. The different outer 
boundary conditions do affect the stellar evolution.
For the case where the photospheric boundary condition 
is adopted at $M_* =10~M_{\odot}$ (MD3e1-HCm10), for example,
the star initially contracts after the boundary
condition is switched at $M_* = 10~M_\odot$, and then 
abruptly inflates at $M_* \simeq 45~M_\odot$. 
The stellar radius exceeds $10^3~R_\odot$ and 
gradually increases with the stellar mass thereafter.
In spite of the different behaviors in the early stages,
the subsequent evolution for $M_* \gtrsim 100~M_\odot$ is
quite similar to that in the case 
with the shock boundary condition throughout (MD3e1). 
The evolution when the boundary condition switching
occurs at $M_* =50~M_{\odot}$ (MD3e1-HCm50)
is much closer to that in the shock-boundary case. 
The uniqueness of the mass-radius relation for 
$M_* \gtrsim 100~M_\odot$ can be explained by the fact that
the argument leading to the analytic expression,  
equation (\ref{eq:rst_analytic}), does not assume a specific 
boundary condition. When the boundary condition is 
switched at $M_* =10~M_{\odot}$ (MD3e1-HCm10),
the stellar interior temperature is higher and thus 
the opacity in the stellar interior ($\propto T^{-3.5}$ according 
to Kramars's law) is lower than for the shock-boundary case (MD3e1)
at the same stellar mass. 
As a result the star begins to release 
its internal energy earlier than for the shock-boundary case. 
Indeed, the timescale equality between $t_{\rm KH}$ and $t_{\rm acc}$  
occurs at $M_* \simeq 40~M_\odot$, earlier than for the shock-boundary 
case, which occurs at the time of abrupt expansion of the stellar radius.

\section{Summary and Discussions}
\label{sec:sum}

We have studied the evolution of stars growing via very
rapid mass accretion with 
$10^{-2}~\msunyr \lesssim \mdot \lesssim 1~\msunyr$, which 
potentially leads to formation of SMBHs in the early universe.
In contrast to previous attempts to address this problem, we
study the stars' evolution by numerically solving the
stellar structure equations 
including mass accretion. 
Our calculations show that stellar evolution in such cases
is qualitatively different from that expected for normal Pop III star
formation, which proceeds at much lower accretion rates 
$\sim 10^{-3}-10^{-2}~\msunyr$.
Rapid mass accretion causes the star to inflate;
the stellar radius further increases monotonically with stellar mass 
at least up to $M_* \simeq 10^3~M_\odot$. 
For masses exceeding $\sim 100~M_\odot$, the star consists of  
a contracting radiative core and a bloated surface convective layer. 
The surface layer, which contains only a small fraction of the total
stellar mass, fills out most of the stellar radius. 
The evolution of the stellar radius in this stage follows a unique
mass-radius relation $R_* \propto M_*^{1/2}$, which reaches 
 $\simeq 7000~R_\odot (\simeq 30$~AU) at $M_* = 10^3~M_\odot$, 
in all the cases with $\ga 10^{-2}~\msunyr$. 
Hydrogen burning begins only after the star becomes 
very massive ($M_* \gtrsim 100~M_\odot$); its onset 
is shifted toward higher masses for higher
accretion rates. 
With very high accretion rates $\mdot \ga 0.1~\msunyr$,
hydrogen is ignited after the stellar mass exceeds $10^3~M_\odot$.
The stellar radius continues to grow as
$R_* \propto M_*^{1/2}$ even after hydrogen ignition.


In this paper we have focused on the early evolution until
the stellar mass reaches $10^3~M_\odot$. 
The subsequent evolution remains unexplored because of
convergence difficulties with the current numerical codes.
If the star continues to expand following the same mass-radius relation 
(\ref{eq:rst_analytic}) also for $M_* > 10^3~M_\odot$, 
the stellar radius at $10^5~M_\odot$ would be $\simeq 400$~AU.
Since the stellar effective temperature remains $\simeq 5000$~K, 
the star hardly emits ionizing photons during accretion.
Therefore, it is unlikely that stellar growth is limited by 
the radiative feedback via formation of an HII region 
as discussed by \citet{HOYY11}.
\citet{Johnson11} also reached an analogous conclusion that 
UV feedback does not hinder SMS formation.  
In their argument, however, the star is assumed to reach the ZAMS and 
to emit a copious amount of ionizing photons, but the expansion 
of the HII region is squelched by rapid spherical inflow.
They also expected that, as a result of confinement of the HII
region, strong emission lines reprocessed from the ionizing photons
(e.g., Ly $\alpha$ and He II) would escape from the accretion envelope
to be an observational signature of these objects.
By contrast, our calculations show that the stellar UV luminosity and thus 
the luminosities in those lines should be much weaker than supposed. 
Note that the argument by \citet{Johnson11} assumes perfect spherical
symmetry, which allows the HII region to be
confined within the accretion envelope.
Given that mass accretion will likely occur through 
a circumstellar disk, the HII region should grow toward the polar 
region where the gas density is much lower than the spherical 
accretion flow \citep[e.g.,][]{HOYY11}. 
This should be the case with the high stellar UV luminosity 
assumed in \citet{Johnson11}.

Even without stellar radiative feedback, stellar growth 
via mass accretion might be hindered by some other process, 
e.g., rapid mass loss. Indeed, evolved massive stars in the Galaxy 
($M_* \sim 10 - 100~M_\odot$), which have 
large radii ($R_* \gtrsim 100~R_\odot$) and high luminosities 
close to the Eddington limit ($L_* \simeq 10^6~L_\odot$), 
generally have strong stellar winds with mass losses 
$\sim 10^{-4}~\msunyr$ \citep[e.g.,][]{HD94}.
Although the line-driven winds of primordial stars are predicted to 
be weak or non-existent \citep{KK06}, 
pulsational instability of massive stars has also been found to 
drive mass loss \citep{BHW01,SU11}.
Further work is necessary to address how massive SMSs could 
form via mass accretion in spite of such disruptive effects.


Stellar evolution under conditions of very rapid mass accretion
as presented and discussed here is mostly relevant to the formation
of stars in the atomic-cooling halos.
However, our results could be also important for 
normal Pop III star formation where H$_2$ molecular cooling
operates. The typical mass accretion rate for this case is around
$10^{-3}~\msunyr$, but in some exceptional situations, 
e.g., when a progenitor cloud core is extremely slow rotating, higher
accretion rates $\mdot \sim 10^{-2}~\msunyr$ can be 
realized \citep[e.g.,][]{HOYY11}.  
Since the stellar effective temperature is low at $\simeq 5000$~K with 
rapid mass accretion, formation of the HII region would be postponed 
until the mass accretion rate falls below $10^{-2}~\msunyr$. 
This would help the primordial star to grow to more than 
$100~M_\odot$ in the molecular-cooling halos 
\citep[see also][]{OP03}.

{\acknowledgements 
The authors thank Francesco Palla, Neal Turner, Rolf Kuiper, 
Kohei Inayoshi, and Naoki Yoshida for fruitful discussions and comments.
T.H. appreciates the support by Fellowship of the Japan
Society for the Promotion of Science for Research Abroad.
K.O. is supported by the Grants-in-Aid by the Ministry of 
Education, Science and Culture of Japan (2168407 and 21244021).
Portions of this work were conducted at the Jet Propulsion Laboratory,
California Institute of Technology, operating under a contract with 
the National Aeronautics and Space Administration (NASA).
}

\bibliography{ms}
\bibliographystyle{apj}

\end{document}